\documentclass[aps,prl,twocolumn,groupedaddress,showpacs,10pt]{revtex4}

\usepackage{amsmath, amssymb, amsthm, mathrsfs, mathtools, amscd}
\usepackage{graphicx, longtable, afterpage, float}
\usepackage{xspace}
\usepackage[latin1]{inputenc}
\usepackage{color}

\newcommand\bbC{\mathbb{C}}

\newcommand\mc[1]{\mathcal{#1}}
\newcommand\on[1]{\operatorname{#1}}
\newcommand\cH{\mc{H}}
\newcommand\hP{\hat P}
\newcommand\hQ{\hat Q}
\newcommand\hA{\hat A}
\newcommand\hU{\hat U}
\newcommand\cC{\mc{C}}
\newcommand\tr{\on{tr}}

\newcommand\ra{\rightarrow}
\newcommand\da{\downarrow}
\newcommand\lra{\longrightarrow}
\newcommand\lmt{\longmapsto}
\newcommand\ld{\lambda}
\newcommand\trho{\tilde{\rho}}
\newcommand\ket[1]{|{#1}\rangle}
\newcommand\bra[1]{\langle{#1|}}
\newcommand\PH{\mc P(\cH)}

\newcommand\eq[1]{(\ref{#1})}

\begin{document}


\title{Contextual Entropy and Reconstruction of Quantum States}


\author{Carmen Maria Constantin, Andreas D\"oring}
\affiliation{Quantum Group, Department of Computer Science, Oxford University, Oxford}


\date{\today}

\begin{abstract}
We introduce a new notion of entropy for quantum states, called contextual entropy, and show how it unifies Shannon and von Neumann entropy. The main result is that from the knowledge of the contextual entropy of a quantum state of a finite-dimensional system, one can reconstruct the quantum state, i.e., the density matrix, if the Hilbert space is of dimension $3$ or greater. We present an explicit algorithm for this state reconstruction and relate our result to Gleason's theorem.
\end{abstract}

\pacs{03.67.-a}

\maketitle



\section{\label{Sec_Intro}Introduction}			
\vspace{-0.2cm}
Quantum Information Theory has brought the importance of information-theoretic concepts to the forefront of physics. Quantum systems have been shown to be able to perform information-theoretic tasks beyond the capabilities of classical systems, e.g. secure key commitment \cite{BB84}, quantum teleportation \cite{BB93}, factoring primes in polynomial time \cite{Sho94}, and many others. Here, we will compare quantum systems to classical systems at the level of their associated information-theoretic notions, in particular we consider \emph{entropy} of physical states. For classical states, Shannon entropy \cite{Sha48} is typically used, for quantum states, von Neumann entropy \cite{vN55}. We will show how a new notion of contextual entropy unifies these two.

Let $\rho$ be a quantum state. In the following, we will distinguish between the (fixed and basis-independent) state $\rho$ itself and the basis-dependent density matrix $\trho$ representing it. The von Neumann entropy $S(\rho)$ of $\rho$ is the Shannon entropy of a specific probability distribution, given by the diagonal elements of $\trho$ in a basis in which $\trho$ is diagonal. Picking such an orthonormal basis means choosing a particular measurement context. Yet, infinitely many other measurement contexts (in which $\trho$ typically is not a diagonal matrix) are available and have a well-defined operational meaning.

In this article, we show that it is fruitful to take all possible measurement contexts into account and consider a family of Shannon entropies, one for each context. This leads to the new notion of the contextual entropy of a quantum state, which is a real-valued function from which the density matrix $\trho$ and hence the state $\rho$ can be reconstructed. This also provides an extension of Gleason's theorem.
\vspace{-0.4cm}

\section{Definition of contextual entropy}
\vspace{-0.2cm}
Consider a finite-dimensional quantum system with Hilbert space $\cH=\bbC^n$, for example a spin system, or a system of $m$ qubits, in which case $n=2^m$. Let $(\hP_1,\ldots\hP_k)$ be a family of projection operators on $\cH$ such that $\hP_i\hP_j=\delta_{ij}\hP_i$ and $\sum_{i=1}^k \hP_i=\hat 1$. Such a family is called a \emph{context} (or sometimes a \emph{resolution of the identity}) and describes a measurement with $k$ outcomes. Conceptually, a measurement context is a `classical perspective' on the quantum system. A context $C:=(\hP_1,\ldots\hP_k)$ can be interpreted as the family of projections onto the eigenspaces of an observable of the form $\hA=\sum_{i=1}^k a_i\hP_i$ with $k$ distinct real eigenvalues $a_i$. The $a_i$ are not determined by the $\hP_i$ (and their actual numerical values do not matter).

Given a context $C=(\hP_1,\ldots\hP_k)$, we can obtain coarse-grained contexts by introducing degeneracy: for example, $C'=(\hP_1+\hP_2,\hP_3,\ldots,\hP_k)$ is another context in which the outcomes $1$ and $2$ cannot be distinguished. We write $C'\leq C$ if each projection in $C'$ is either in $C$ or is the sum of projections in $C$, and every projection in $C$ shows up in such a sum exactly once. In this case we say that the context $C'$ is \emph{coarser} (or \emph{more degenerate}) than the context $C$. This gives a partial order on the set $\cC$ of all contexts of our quantum system.

A context $C$ is called \emph{maximal} if there are no contexts $\tilde C\neq C$ such that $C\leq\tilde C$. If $C=(\hP_1,\ldots,\hP_n)$ is maximal, then the $n$ projections $\hP_i$ are of rank $1$ and $C$ corresponds to a non-degenerate measurement. Each rank-$1$ projection $\hP_i$ determines a unit vector $\ket{\psi_i}$ such that $\ket{\psi_i}\bra{\psi_i}=\hP_i$. This vector is unique up to a phase. Hence, with each maximal context we can associate an orthonormal basis $(\ket{\psi_1},\ldots,\ket{\psi_n})$ of $\cH$ that is unique up to phases. If $C=(\hP_1,\ldots,\hP_k)$ with $k<n$ is a non-maximal context, one can still find a (non-unique) orthonormal basis of $\cH$ such that all the $\hP_i$ in $C$ are given by diagonal matrices with respect to this basis, since all the projections in $C$ commute.
 
Not all contexts $C,\tilde C\in\cC$ can be compared with respect to the order defined above. This corresponds to the fact that some measurements are incompatible. However, incompatible contexts can be coarse-grained to the same context $C$. Consider for example the three spin operators $\hat S_i$ in $\cH=\bbC^2$. Each $\hat S_i$ defines a context $C_i$, given by the two projections onto the eigenspaces of $\hat S_i$. As is well-known, the $\hat S_i$ do not commute, but they all commute with the total spin $\hat{\mathbf{S}}^2$. The latter defines a context $C$ that just contains the identity operator $\hat 1$ and hence is coarser than $C_1,C_2$ and $C_3$. This indicates that the partial order gives a rich structure to the set $\cC$ of all contexts. For some results on the structure of $\cC$ see \cite{HD10,DB11}.

Let $\rho$ be a state of a finite-dimensional quantum system, described by a density matrix $\trho$. We will say that $\rho$ is \emph{diagonal in a context} $C=(\hP_1,\ldots,\hP_k)$ if $\rho$ is of the form $\rho=\sum_{i=1}^k \ld_i\hP_i$. This is justified by the following: if $\mc B=(\ket{\psi_1},\ldots,\ket{\psi_n})$ is an orthonormal basis of $\cH$ such that all projections $\hP_i$ in $C$ are given by diagonal matrices with respect to $\mc B$, then $\trho$ is diagonal with respect to $\mc B$ if and only if $\rho=\sum_i \ld_i\hP_i$.

Given a measurement with $k$ outcomes described by a context $C$, the quantum state $\rho$ assigns a probability $p_i:=\on{Prob}(i;\rho)=\tr(\trho\hP_i)$ to each outcome $i=1,\ldots,k$.
This defines a probability distribution $P_C:=(p_1,\ldots,p_k)$. Since a context can be interpreted as a \emph{classical perspective} on a quantum system, it makes sense to assign a (classical) Shannon entropy to the probability distribution $P_C$, which is $H(P_C)=-\sum_{i=1}^k p_i\ln p_i$. Yet, no context $C$ is preferred over the others, so we consider \emph{all} contexts $C\in\cC$. This gives a map
\begin{align*}
			E_{\rho}:\cC &\lra [0,\ln n]\\			\nonumber
			C=(\hP_1,\ldots,\hP_k) &\lmt H(P_C)=H(\tr(\trho\hP_1),\ldots,\tr(\trho\hP_k)).
\end{align*}
Here, $[0,\ln n]$ denotes the real interval from $0$ to $\ln n$. The map $E_{\rho}$ is called the \emph{contextual entropy} of the quantum state $\rho$. It is a family of Shannon entropies, one for each context $C\in\cC$.
\vspace{-0.4cm}
\section{Properties of contextual entropy}
\vspace{-0.2cm}
We show two important properties of contextual entropy. The first of these highlights the connection between the von Neumann and the contextual entropy of a state and will be useful for our reconstruction algorithm. The second one shows that the contextual entropy is a monotone map.

\textbf{Von Neumann entropy from contextual entropy.} We first show that there exists at least one maximal context $C_{\rho}\in\cC$ such that $E_{\rho}(C_{\rho})=S(\rho)$, the von Neumann entropy of the state $\rho$: let $C_{\rho}=(\hP_1,\ldots,\hP_n)$ be a maximal context in which $\rho$ is diagonal, that is, $\rho=\sum_{i=1}^n \ld_i\hP_i$. Then $P_{C_{\rho}}=(\tr(\trho\hP_1),\ldots,\tr(\trho\hP_n))= (\ld_1,\ldots,\ld_n)$ and hence $E_{\rho}(C_{\rho})=H(P_{ C_{\rho}})=-\sum_i \ld_i\ln\ld_i=S(\rho)$, which is what we wanted to show.

Secondly, recall that a vector $\mathbf{r}=(r_1,\ldots,r_n)$ of real numbers is \emph{majorised} by another real vector $\mathbf{s}=(s_1,\ldots,s_n)$ if $\sum_{i=1}^n r_i=\sum_{i=1}^n s_i$ and, for all $k<n$, $\sum_{i=1}^k r_i^\da \leq \sum_{i=1}^k s_i^\da$, where the $r_i^\da$ are the components of $\mathbf{r}$ rearranged in decreasing order, and similarly for the $s_i^\da$ \cite{Bha97}.

Using the majorisation order, we can show a stronger result: the von Neumann entropy $S(\rho)$ of a quantum state $\rho$ is the \emph{minimal} value of the contextual entropy $E_{\rho}(C)$ when $C$ is varying over maximal contexts.

To see this, let $C=(\hP_1,\ldots,\hP_n)$ be any maximal context. Necessarily, all projections $\hP_i\in C$ are of rank $1$. Consider the matrix representation of $\hP_i$ with respect to the orthonormal basis $\mc B_C=(\ket{\psi_1},\ldots,\ket{\psi_n})$ associated with the context $C$. This matrix has a single $1$ in position $(i,i)$ and zeros everywhere else. Then $p_i=\tr(\trho\hP_i)=\tr(\trho\hP_i^2)=\tr(\hP_i\trho\hP_i)$ is the $i$th diagonal element of the density matrix $\trho$ when written with respect to the basis $\mathcal{B}_C$. Hence, the probability distribution $P_C=(p_1,\ldots,p_n)$ in context $C$ determined by the state $\rho$ consists of the diagonal elements of the density matrix $\trho$ when the latter is written with respect to the basis $\mathcal{B}_C$ associated with $C$.

The Schur-Horn theorem \cite{Hor54} states that the vector $\mathbf{l}=(\ld_1,\ldots,\ld_n)$ of eigenvalues of a Hermitian diagonal matrix $M$ majorises the vector $\mathbf{k}=(\kappa_1,\ldots,\kappa_n)$ of diagonal elements of the Hermitian matrix $\hU M\hU^{-1}$ obtained after a change of basis. In our situation, $M=\trho$, and a change of basis amounts to a change of maximal context. So, if $C_{\rho}=(\hQ_1,\ldots,\hQ_n)$ is a maximal context in which $\rho$ is diagonal and $C$ is any other maximal context, then the vector $P_{C_{\rho}}=(\ld_1,\ldots,\ld_n)$ of eigenvalues of $\rho$ majorises the vector $P_{C}=(p_1,\ldots,p_n)$ of diagonal elements of $\trho$ written with respect to the basis associated with $C$.

It is well-known that the Shannon entropy reverses the majorisation order, so it follows that the contextual entropy $E_{\rho}$ takes its minimal value on the set of maximal contexts of the form $C_{\rho}$ for which $\trho$ is a diagonal matrix. We saw above that for such contexts $C_{\rho}$, the value $E_{\rho}(C_{\rho})$ of the contextual entropy is the von Neumann entropy $S(\rho)$ of the quantum state. This completes the proof.

\textbf{Monotonicity.} Now consider two contexts $C,C'\in\cC$ such that $C'$ is coarser than $C$. We want to compare $E_\rho(C')$ and $E_\rho(C)$. Let $C'=(\hP_1,\ldots,\hP_k)$, and let $C=(\hQ^1_1,\ldots,\hQ_{l_1}^1,\hQ^2_1,\ldots,\hQ^2_{l_2},\ldots,\hQ^k_1,\ldots,\hQ^k_{l_k})$, where $\sum_{j=1}^{l_i} \hQ^i_j = \hP_i$ for all $i=1,\ldots,k$. Given a quantum state $\rho$, we have $p_i=\tr(\trho\hP_i)=\tr(\trho\hQ^i_1)+\ldots+\tr(\trho\hQ^i_{l_i})$ for all $i=1,\ldots,k$. Let us denote $q^i_j:=\tr(\trho\hQ^i_j)$, where $i=1,\ldots,k$ and $j=1,\ldots,l_i$. Then, using the recursion property of Shannon entropy (see e.g. \cite{BZ06}),
\begin{align*}
			E_{\rho}(C) = &\; H(q^1_1,\ldots,q^1_{l_1},q^2_{1},\ldots,q^2_{l_2},\ldots,q^k_1,\ldots,q^k_{l_k})\\	\nonumber
			= &\;H(p_1,\ldots,p_k) + \sum_{i=1}^k p_i H\left(\frac{q^i_1}{p_i},\ldots,\frac{q^i_{l_i}}{p_i}\right)\\
			= &\;E_{\rho}(C')+\sum_{i=1}^k p_i H\left(\frac{q^i_1}{p_i},\ldots,\frac{q^i_{l_i}}{p_i}\right).
\end{align*}
Since the last term above is always non-negative, it follows that $E_{\rho}(C')\leq E_{\rho}(C)$ for all contexts $C'\leq C$. Hence, the map $E_{\rho}:\cC\ra [0,\ln n]$ is indeed order-preserving.
\vspace{-0.4cm}

\section{\label{Sec_Reconstruction}Reconstruction of quantum states}
\vspace{-0.2cm}

We now show that the contextual entropy of a quantum state $\rho$ contains enough information to uniquely reconstruct the density matrix $\trho$ if the dimension of the Hilbert space is at least $3$. We assume that we can identify a maximal context $C_\rho$ for which $E_\rho$ takes its minimal value among all maximal contexts. (In general, the minimal value of $E_{\rho}$ will be attained in many different maximal contexts, but any of them will do.) We remark in passing that the same reconstruction algorithm would also work for contextual R\'enyi entropies.

\textbf{Pure states.} Assume that $\rho$ is pure, i.e., $\rho=\ket\psi\bra\psi$ for some unit vector $\ket\psi\in\cH$. We want to determine $\ket\psi\bra\psi$ from the contextual entropy $E_{\rho}$.

We first note that a state $\rho$ is pure if and only if $E_{\rho}(C_{\rho})=0$. This follows from the previous section, where we showed that $E_{\rho}(C_{\rho})=S(\rho)$, the von Neumann entropy of $\rho$. The latter is equal to $0$ if and only if $\rho$ is a pure state.

Let $C=(\hP_1,\ldots,\hP_n)$ be a context for which $E_{\rho}(C)=0$. Then there is a unique $\hP_{i_0}\in C$ such that $\ket\psi\bra\psi=\hP_{i_0}$. Using finitely many values of the contextual entropy $E_\rho$, it is possible to determine which of the $n$ projections $\hP_1,\ldots,\hP_n$ equals $\ket\psi\bra\psi$. For this, consider $n$ unitary matrices $\hU_1,\ldots,\hU_n$ such that $\hU_i\hP_i\hU_i^{-1}=\hP_i$ and, for all $1\leq j\leq n$, $j\neq i$,
\begin{equation*}
			\hU_i\hP_j\hU_i^{-1}\notin\{\hP_1,\ldots,\hP_n\}.
\end{equation*}
That is, $\hU_i$ leaves the projection $\hP_i$ invariant and `rotates' the other projections $\hP_j$, $j\neq i$ without resulting in a permutation of any of them. Such unitaries always exist if $\dim\cH\geq 3$: for example, a unitary keeping $\hP_1$ fixed and rotating all other $\hP_j$ has a matrix of the form
\[
			\hU= \left(\begin{array}{c|c} 1&\textbf{0}^T\\\hline \textbf{0}&\hU'\end{array} \right),
\]
where the unitary $\hU'\in\mc U(n-1)$ can be taken to be a rotation by some small angle around an axis different from all the $n-1$ coordinate directions for $j=2,\ldots,n$.

Consider the maximal contexts of the form $C_i:=(\hU_i\hP_1\hU_i^{-1},\ldots,\hU_i\hP_n\hU_i^{-1})$ for $i=1,\ldots,n$. Only the context $C_{i_0}$ contains the projection $\hP_{i_0}=\ket\psi\bra\psi$, which is the state that we are looking for. Hence, the density matrix $\trho$ is only diagonal with respect to the orthonormal basis associated with the maximal context $C_{i_0}$, while it is not diagonal with respect to the bases associated with any of the contexts $C_j$, $j\neq i_0$. If the density matrix $\trho$ of a projection $\ket\psi\bra\psi$ is not diagonal with respect to some basis, then there are at least two non-zero entries on the diagonal, so the probability distribution $(p_1,\ldots,p_n)$ given by the diagonal elements of $\trho$ has Shannon entropy strictly larger than $0$.

Hence, $E_{\rho}(C_{i_0})=0$, while $E_{\rho}(C_j)>0$ for all $j\neq i_0$. In this way, we can identify $i_0$ and can determine the state $\ket\psi\bra\psi$.

\textbf{Mixed states.} We now show how to reconstruct $\trho$ from $E_{\rho}$ when $\rho$ is a mixed state and $\dim\cH\geq 3$.

\textbf{Step 1.} Find a maximal context $C_{\rho}=(\hP_1,\ldots,\hP_n)$ for which the value $E_{\rho}(C_\rho)$ is minimal (but larger than $0$ since $\rho$ is mixed). The Schur-Horn theorem implies that the state which we want to determine is diagonal in $C_{\rho}$, that is, $\rho=\sum_{i=1}^n \ld_i\hP_i$. We have to find the eigenvalues $\ld_i$ of $\rho$.

\textbf{Step 2.} Let $C_i:=(\hP_i,\hat 1-\hP_i)$ be the context containing just $\hP_i$ and its complement, for each $i=1,\ldots,n$. Then $E_\rho(C_i)$ is the Shannon entropy of the probability distribution $(\tr(\trho\hP_i),1-\tr(\trho\hP_i))$. 

Of course, the actual values $\ld_i:=\tr(\trho\hP_i)$ and $1-\ld_i=1-\tr(\trho\hP_i))$ are unknown so far. We just have the binary entropy
\begin{equation*}
			E_\rho(C_i)=-x_i\ln x_i-(1-x_i)\ln(1-x_i).
\end{equation*}
Solving this, we obtain two solutions, $c_i$ and $1-c_i$. We can assume without loss of generality that $c_i\leq\frac{1}{2}\leq 1-c_i$. For each $i$, the eigenvalue $\ld_i$ of the density matrix $\trho$ is either $c_i$ or $1-c_i$.

\textbf{Step 3.} We write the solutions $c_1\ldots,c_n$ and $1-c_1,\ldots,1-c_n$ in two rows such that $1-c_i$ is beneath $c_i$. In order to find the eigenvalues $\ld_i$ of the density matrix $\trho=\sum_{i=1}^n \ld_i\hP_i$, written with respect to the basis associated with $C_\rho$, we must choose $n$ numbers from this table, one from each column, such that their sum is equal to $1$ (since $\sum_{i=1}^n \ld_i=1$). At least one such solution must exist, since we assumed that $E_\rho$ is the contextual entropy of some quantum state $\rho$.

\textbf{(a)} If the numbers $c_i$ in the top row add up to $1$, then we have found the unique solution: $\ld_i:=c_i$ for each $i=1,\ldots,n$. Picking any $1-c_i$ instead of $c_i$ would make the total sum greater than $1$.

\textbf{(b)} If the sum $S:=\sum_{i=1}^n c_i$ of the elements in the top row is smaller than $1$, then we must replace at least one value $c_j$ from the top row with $1-c_j$ from the bottom row. Note, however, that since the entries of the bottom row are all greater or equal to $\frac{1}{2}$, we can only pick exactly one such entry, since picking two or more would make the sum of our $n$ chosen elements greater than $1$.

To determine which $c_j$ (from the top row) must be replaced by $1-c_j$ (from the bottom row) among the $n$ elements we pick, we note the following: the sum $c_1+\ldots+c_{j-1}+(1-c_j)+c_{j+1}+\ldots+c_n$ must equal $1$, which implies that $c_j$ must have the value $c:=\frac{(c_1+\ldots+c_n)}{2}=\frac{S}{2}<\frac{1}{2}$. 

\textbf{(b1)} If the value $c$ appears only once among the entries of the top row, for example in the $j$th column, then our unique solution is $\ld_j=1-c_j$ and $\ld_i=c_i$ for all $i\neq j$.

\textbf{(b2)} If, however, the value $c$ appears twice among the entries of the top row, the bottom row has $1-c$ in the corresponding two columns. This implies that the top row has $0$s in all other columns, since otherwise the sum of $1-c$ (the unique entry picked from the bottom row) and the sum of the $n-1$ entries from the other colums of the top row would be larger than $1$. This also implies that $c$ cannot appear three or more times among the entries of the top row. Assume that the two entries of $c$ appear in the $j$th and $k$th columns. Our state is then either
\vskip -12pt
\begin{equation}			\label{r1}
			\rho=c\hP_j+(1-c)\hP_k
\end{equation}
\vskip -10pt
\noindent or
\vskip -20pt
\begin{equation}			\label{r2}
			\rho=c\hP_k+(1-c)\hP_j.
\end{equation}
In order to determine which of these is the correct solution, consider a unitary $\hU$ which rotates, but does not permute, all the projections $\hP_1,\ldots,\hP_n$ except for $\hP_j$, which it leaves unchanged. Here, we need $\dim\cH\geq 3$. The $j$th eigenvalues of $\hU^{-1}\rho\hU$ and $\rho$ (that is, the eigenvalues for the joint eigenvector determined by $\hP_j$) coincide and are equal to $\ld_j$, while the other eigenvalues are distinct in general. We consider the contexts of the form $W_i=(\hU\hP_i\hU^{-1},\hat 1-\hU\hP_i\hU^{-1})$ and solve the equations
\vskip -16pt
\begin{equation*}
			E_\rho(W_i)=-d_i\ln d_i- (1-d_i)\ln (1-d_i).
\end{equation*}
\vskip -4pt
We write the solutions in a second table, again with the convention that the top row contains entries smaller than $\frac{1}{2}$. We then repeat the procedure detailed above (from \textbf{Step 3} onwards)  of choosing $n$ numbers adding up to $1$, this time from the second table. These numbers will be equal to the diagonal entries of the matrix $\hU^{-1}\trho\hU$, written in the basis in which $\trho$ is diagonal. 

Because of our choice of unitary, the diagonal entries of the density matrix $\hU^{-1}\trho\hU$ will contain at least three non-zero elements, so we do not enter the \textbf{(b2)} branch of our algorithm again, since the top row of the second table will also contain at least three non-zero entries. Hence, this time there will be a unique choice of $n$ entries adding up to $1$. In particular, the $j$th element of this solution equals the $j$th diagonal entry of the matrix $\hU^{-1}\trho\hU$, which is the same as $\ld_j$, the $j$th eigenvalue of $\rho$. This allows us to choose the correct quantum state from the two possible solutions \eq{r1} and \eq{r2}. We finally remark that for $\dim\cH=2$, the qubit case, we can determine the state up to the ambiguity between \eq{r1} and \eq{r2}.
\vspace{-0.4cm}

\section{Relation to Gleason's theorem}
\vspace{-0.2cm}
Let $\cH=\bbC^n$, $n\geq 3$, and let $\mu$ be a (finitely additive) probability measure $\mu:\PH\ra [0,1]$ on the projections of $\cH$, that is $\mu(\hat 1)=1$ and if $\hP\hQ=\hQ\hP=\hat 0$, then $\mu(\hP+\hQ)=\mu(\hP)+\mu(\hQ)$.
 
Gleason's theorem \cite{Gle57} (which we only consider for the finite-dimensional case here) states that for every such probability measure $\mu$, there exists a quantum state $\rho_\mu$ such that, for all projections $\hP\in\PH$,
\vskip-15pt
\begin{equation}			\label{Eq_Gleason}
			\tr(\trho_\mu\hP)=\mu(\hP).
\end{equation}
\vskip-4pt
Conversely, every quantum state $\rho$ gives a probability measure $\mu_\rho:\PH\ra [0,1]$ simply by setting $\mu_\rho(\hP):=\tr(\trho\hP)$ for all $\hP\in\PH$.

A probability measure $\mu$ defines a probability distribution $P_C=(\mu(\hP_1),\ldots,\mu(\hP_k))$ for each context $C=(\hP_1,\ldots,\hP_k)$ . Gleason's theorem shows that if we have a family $(P_C)_{C\in\cC}$ of probability distributions, one for each context, that come from a measure $\mu$, then there is a unique quantum state $\rho_\mu$ such that eq. \eq{Eq_Gleason} holds.

Given $\mu$, we can construct the corresponding contextual entropy $E_\mu$: to each context $C=(\hP_1,\ldots,\hP_k)$, we assign the Shannon entropy of the probability distribution $P_C$. As was shown in the previous section, one can reconstruct the quantum state $\rho_\mu$ from its contextual entropy $E_\mu$. This requires a \emph{single} real number $E_\mu(C)$ for each context $C=(\hP_1,\ldots,\hP_k)$ instead of the $k$ numbers $\mu(\hP_1),\ldots,\mu(\hP_k)$. Moreover, we obtain an explicit density matrix from our reconstruction, while Gleason's theorem merely shows that a density matrix must exist. In this sense, our approach via the contextual entropy is an extension of Gleason's result.

On the other hand, Gleason's theorem guarantees that \emph{every} probability measure $\mu:\PH\ra [0,1]$ corresponds to a quantum state, while we had to assume that the map $E_\rho:\cC\ra [0,\ln n]$ which we use in our reconstruction actually is the contextual entropy of some quantum state. 
\vspace{-20pt}

\section{Summary and outlook}
\vspace{-8pt}
Given a quantum state $\rho$ on a finite-dimensional Hilbert space $\cH$ and a measurement context $C=(\hP_1,\ldots,\hP_n)$, we can extract the probability distribution $P_C=(\tr(\trho\hP_1),\ldots,\tr(\trho\hP_n))$ by repeated preparations and measurements. In contrast to the quantum state itself, measurement contexts have direct operational meaning. The contextual entropy $E_\rho:\cC\ra [0,\ln n]$, which is a real-valued function, assigns to each probability distribution $P_C$ its Shannon entropy and hence encodes data that can be extracted operationally from the quantum state $\rho$.

The fact that the state $\rho$ can be reconstructed from its contextual entropy $E_\rho$ if $\dim\cH\geq 3$ provides a new, information-theoretic characterisation of quantum states that takes contextuality into account explicitly. The results in this article connect directly with the so-called \emph{topos approach} to quantum theory \cite{DI07,D07} in which contextuality is a key concept. In future work, we will develop these connections in depth and will also consider infinite-dimensional systems.

We presented a number of properties of contextual entropy and discussed how the reconstruction of a quantum state from its contextual entropy relates to Gleason's theorem. As matters stand, the properties we presented do not characterise contextual entropy fully: there are functions $F:\cC\ra [0,\ln n]$ that have all the properties we discussed, but are not the contextual entropy of any quantum state.

Finding an axiomatic characterisation of exactly those functions which are contextual entropies of quantum states would, together with our reconstruction algorithm, provide an alternative proof of Gleason's theorem. This is an interesting and non-trivial open problem.

\textbf{Acknowledgements.} We thank Oscar Dahlsten, Andrei Constantin, Daniel Marsden, Rui Soares Barbosa and Chris Isham for discussions and suggestions, and we thank Samson Abramsky and Bob Coecke for support. C.M.C. is supported by an EPSRC graduate scholarship.

{}

\end{document}